\documentclass[journal,comsoc]{IEEEtran}
\usepackage{graphicx,color} 
\usepackage{url}

\usepackage[bookmarks=false]{hyperref}
\hypersetup{%
         pdfdisplaydoctitle,%
         colorlinks=true,%
         allcolors=black,%
         pdfstartview=Fit,%
   }%
\usepackage{multicol}
\usepackage{orcidlink}
\newcommand*{\orcid}[1]{\vspace{-5pt}\orcidlink{#1}}%
\title{Conceptualizing Trustworthiness and Trust \\in Communications}
\author{Gerhard P. Fettweis\textsuperscript{1,2}\orcid{0000-0003-4622-1311} \IEEEmembership{(Fellow, IEEE)}, Patricia Grünberg\textsuperscript{1}\orcid{0009-0002-1342-8447}, Tim Hentschel\textsuperscript{1}\orcid{0009-0009-8698-0874},\vskip0.5cm Stefan Köpsell\textsuperscript{1,3}\orcid{0000-0002-0466-562X}
\IEEEmembership{(Member, IEEE)}\vskip0.1cm
\footnotesize{\textsuperscript{1}Barkhausen Institut}\vskip0.01cm
\footnotesize{\textsuperscript{2}Vodafone Chair TU Dresden, Centres 6G-life \& CeTI \& 5G\textsuperscript{++}Lab Germany \& Else Kröner Fresenius Center (EKFZ) for Digital Health }\vskip0.01cm
\footnotesize{\textsuperscript{3}Chair of Privacy and Security TU Dresden, Centres CeTI \& TUDiSC}\\
This work has been submitted to the IEEE for possible publication. Copyright may be transferred without notice, after which this version may no longer be accessible.
}
\begin{document}

\maketitle
\begin{abstract}
Trustworthiness and trust are fundamental factors in societies that enable us to interact and enjoy mingling in crowds without fear. As robotic devices start permeating our daily lives, they must behave as completely trustworthy objects so that people will accept them just as they would trust other people when interacting with them in their daily lives.

As trust and trustworthiness have been researched in social sciences for many years, this opens the question: How can we learn from system models and findings from social sciences to translate such learnings into requirements for future technical solutions? This is of particular importance now, as 5G and 6G cellular communications open the door for the Tactile Internet --- connected robotics interacting with humans. We present a novel holistic approach on how to tackle trustworthiness systematically in the context of communications. We propose a first attempt to incorporate objective system properties and subjective beliefs to establish trustworthiness-based trust.
\end{abstract}


\section{Introduction --- A User Perspective}
For over a century, humans have dreamed of personal robotic assistants that would make our lives easier. These robots could take on a variety of tasks, from cleaning our rooms to providing companionship. Hardly any of us can judge whether these robots are ``good guys'' or ``bad guys'' in a future digitization-shaped society. We call these future societies \emph{digitization-shaped}  since the foreseeable proliferation of digital products and services as ``members'' of society will be formative, and we humans will have to trust those digital ``members'' of society. However, rather than blindly trusting them, we should base our trust decisions on trustworthiness, i.e., objectively measurable properties of the system. But when are such systems trustworthy? 6G-enabled systems, as any technical system, can only ``behave'' human-like and hence, can only be trustworthy in a human (and humane) sense if engineers design them to be so. Engineers must build an understanding of this and make systems trustworthy by design. 

It is not far-fetched to assume that with the coming of 6G-cellular-enabled products, we humans will continue our practice of delegating our trust decisions, putting our trust in certain organizations that place a label of trustworthiness on these new products. We will trust those organizations to have the technical ability to assess the very features of a product or service that fundamentally are a proxy for their ``behavior'' in human-machine interaction and ultimately in a future society. 

We are currently in an era where societies increasingly depend on interconnected digital communications as the backbone of our economy, education, government, as well as personal social interactions. However, hardly any of us have the expertise or the tools needed to assess whether our digital communicating world is genuinely reliable in the most literal sense. Even experts often lack access to the comprehensive and detailed information required to fully evaluate a complex product or system. Therefore, when a decision needs to be made, as humans, we simplify the complex problem of understanding all risk factors by trusting most and concentrating on understanding the risk of the crucial variables~\cite{luhmann2017trustandPower}. There are many examples in our daily life. We trust a car when getting in that it is safe; we trust to drive on one side and that other cars stick to the same side; this way we can focus our effort on driving the car safely. Trust is essential for the proper functioning of societies.

How can we translate research findings from social sciences gained on human-human interaction to our challenge of communications in the digital world of today and the future? In particular, how can we possibly build a model of human-digital interaction to ensure that highly interconnected digital technologies incorporate fundamental elements that enable humans to make informed trust decisions?

With 5G and 6G communications, we are at the doorstep of seeing robotic communicating devices enter our world. Lawn mowers, vacuum cleaners, and robo-cars are just the very first examples of the Tactile Internet~\cite{fettweis_tactile_2014}. We need to understand how we can have these robotic devices enter our lives not as untrustworthy oddballs but as trusted elements of our daily physical lives. We will then depend on the trustworthiness of the underlying communication system to build trust for our decision makings. As engineers and computer scientists, we know that successful attacks on digital systems are realistic from a technology perspective. 

To comprehend the research challenge of addressing trustworthiness, we need a model that incorporates all elements and characteristics involved. We, therefore, need to translate knowledge from human-human to human-digital interactions. Using the well-known model of trustworthiness and trust from social sciences as a basis~\cite{mayer_integrative_1995}, this paper proposes a new model as one foundation for building trust and trustworthiness in human-digital interactions. It should serve as a basis for future research on making information and communication technologies and products truly trustworthy.

It should have become apparent that we have used the terms \textit{trust} and \textit{trustworthiness} very deliberately as they are fundamentally different concepts and a large part of this article is based upon that differentiation.

In the next section, we first review the findings of social sciences. Then we look at the technical aspects of trustworthiness and their characteristics, followed by a section on the need to be able to measure. This is the basis for presenting our main contribution, i.e., a new model for trust and trustworthiness in the context of human-digital interactions and communications.
Although (in this early stage of research) we have not conducted our own empirical evaluation, we provide examples from existing literature to underpin our model.
 
\section{A Brief Review of a Social Sciences Model of Trust\label{mark-II.}}
\begin{figure}
    \centering
    \includegraphics[width=\linewidth]{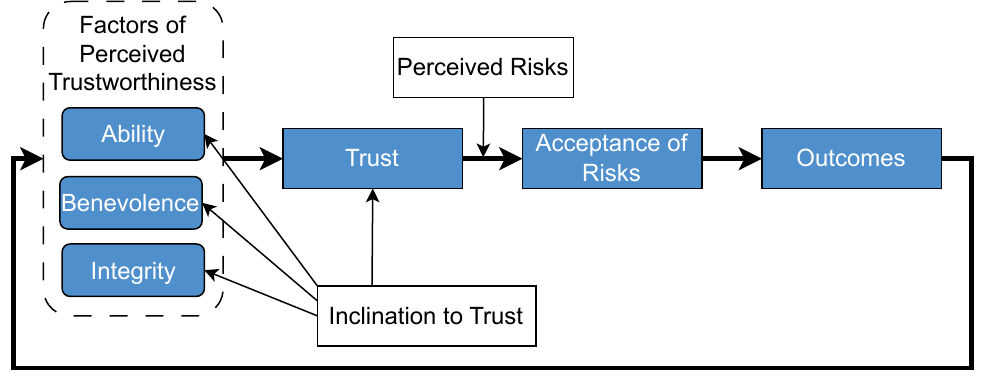}
    \caption{The ABI model of trustworthiness and trust by Mayer et al.~\cite{mayer_integrative_1995}}
    \label{fig:SocioModel}
\end{figure}

Over the last few decades, the social sciences and humanities have produced a multitude of definitions and explanations as well as empirical findings on the topic of trust~\cite{grunberg_vertrauen_2013}. Hence, we reflect on the insights from the humanities and social sciences in the area of trust and trustworthiness before conceptualizing them for the area of communications engineering. 

On a micro level, both psychology and educational science predominantly explore \textit{interpersonal trust} in specific interaction situations. However, this micro perspective is not sufficient to convey the multiple facets of trust. A macro-level approach takes place primarily in sociology and political science, with a particular focus on \textit{system trust}. Political science studies the concept of trust mainly from a developmental dynamic perspective, aiming to explain the development of \textit{trust in democracy or institutions}. Economic science studies the concept of trust both from a basic theory and a problem-oriented perspective, the latter primarily in organizational and marketing research. Specifically, organizational research focuses on questions of what \textit{organizational trust} is, what factors influence it, and what effects it can have on the organization. By contrast, philosophy investigates the connection between trust and morality as well as the relevance of trust for society. In comparison, communication science examines less the question of trust, but rather the credibility of the media. 

The spectrum of social science research on trust is very broad. Still, there has been a lack of consolidation of the research achievements on the topic of trust. Therefore, and certainly owing to the different scientific perspectives and their varying emphases, there is no uniform understanding of the term of \textit{trust}. Yet, the diverse disciplinary interest shows that trust is not a simple concept, but a heterogeneous and complex subject.

To investigate the concept of trust, we use the widely received integrative ABI trust model of Mayer et al. \cite{mayer_integrative_1995} on organizational trust. The focus of this model (Fig.~\ref{fig:SocioModel}), which analyzes the circumstances of trust emergence, is primarily on the trustworthiness of the trusted party as perceived by the trust giver. The authors define trust as ``the willingness of a party to be vulnerable to the actions of another party based on the expectation that the other will perform a particular action important to the trustor, irrespective of the ability to monitor or control that other party''~\cite{mayer_integrative_1995}. Particular emphasis is placed on the moment of vulnerability and the willingness to take a certain risk. The integrative model explains the emergence of trust in a situation in which a relationship between the two parties has not yet evolved. The factors of perceived trustworthiness influence the emergence of trust. However, the expectation of trust does not yet involve any risk, because, according to the model, risk is only the result of trust. This view is in contradiction to the views of other authors, who presume a risk perception that requires trust in the first place.

According to the ABI model, the assessment of trustworthiness as the basis for an act of trust is influenced by the evaluation of the following three factors: ability, integrity, and benevolence. Abilities include skills, competencies, and attributes that enable a person to solve a specific problem. Benevolence describes how much the trusted party considers the well-being of the trusting party in his or her actions. Integrity implies qualities such as honesty and reliability. Although the factors can be considered separately, they are interrelated and can influence each other. 

In the ABI model, trustworthiness is understood as a continuum. At the beginning of a relationship, the trust giver has little to no information about the trust taker. In particular, the assessment of benevolence is difficult. Therefore, Mayer et al. assume that the perception of integrity is key to the initial trust building. As the relationship progresses, both actors obtain more information and gain a better assessment of ability and benevolence. Thus, as the relationship progresses positively, the degree of trust may increase.

The authors themselves reflect the limitations of the integrative trust model: a) the model is only designed to explain trust between two individuals; b) it is unidirectional in design and ignores the development of mutual trust; c) the model was designed to focus on trust in an organizational setting, with the result that generalization to other domains is not immediately possible. Moreover, this is merely a theoretical model and there are no specific assumptions about the dependencies between the individual trust factors.

Although the ABI model is the foundation of much follow-up research on trust and trustworthiness, the number of empirical studies is still limited. A recent one, which empirically supports the ABI model with respect to innovation networks, is described in~\cite{mayerInnovationNetworks}. Moreover, a recent meta-study~\cite{hancock}, which reviewed more than 300 studies, supports the ABI model by extending it with more details related to the trustee, trustor and context-environmental factors.

Parallel to the model of Mayer et al., the theory of public trust was developed in communication science, which also works with different trust factors \cite{bentele_offentliches_1994}. A distinction is made between subject-specific trust factors (factual competence and problem-solving competence), socio-psychological trust factors (social behavior, communication behavior {[}transparency, consistency{]}, character), and socio-normative trust factors (sense of responsibility, ethical behavior). The more completely and intensively the trust factors apply, the more likely are trust gain, trust formation or trust constitution. If the factors apply only partially and/or to a lesser extent, the greater the likelihood of a reduction or loss of trust.

Given these foundations in social sciences, can we come up with a model of trust and trustworthiness that helps explain the trust relationship between humans as trust givers and technology as trust takers?

\section{Trustworthiness and its Measurability from a technical perspective\label{mark-III.}}
If we shift the focus towards the interaction of humans with technical systems, the definition of trust as given in the previous section can basically remain the same, i.e., the willingness of a human to be vulnerable to the actions of a technical system based on the expectation that the system will perform a particular action important to the human, irrespective of the ability to completely monitor or control the technical system. But what about the trustworthiness of the technical system and its measurability?

\subsection{Trustworthiness\label{mark-A1}}
The ISO/IEC standard TS 5723:2022 \cite{isoiec_trustwhorthiness_2022} defines trustworthiness as the ``ability to meet stakeholders’ expectations in a verifiable way''. Similarly, NIST sees trustworthiness as an objective property of a system, since the ``trustworthiness of a system is based on the concept of assurance''~\cite{nist_engineering_2022}. Note that other concepts like ``quality''~\cite{iso_quality_2015}, ``quality of service'', or ``quality of experience'' have a similar broad meaning, according to their definitions. Nevertheless, it makes sense to have different terms, since the wording emphasizes different aspects. For example, quality of service is often related to the fulfillment of non-functional requirements, such as latency or bandwidth. While trustworthiness is often used interchangeably with security and privacy-related aspects, it is important to stress that the fulfillment of expectations of stakeholders of a given system spans a much broader scope of requirements for the design of a truly trustworthy system.

\subsection{Characteristics of Trustworthiness and their measurability\label{mark-B1}}
A recent attempt to map the topic of trustworthiness onto the communications field was published in~\cite{fettweis_6g_2022}. Looking at the topic of the trustworthiness of a communication system, several important indicators were identified, such as:
\begin{multicols}{2}
\begin{itemize}
\item Accountability
\item Authenticity
\item Availability
\item Confidentiality
\item Integrity
\item Privacy
\item Reliability
\item Resilience
\end{itemize}
\end{multicols}
Missing in this list are certainly indicators mentioned in the ISO/IEC standard \cite{isoiec_trustwhorthiness_2022}, e.g.:
\begin{multicols}{3}
\begin{itemize}
\item Accuracy
\item Controllability
\item Robustness
\item Safety
\item Transparency
\item Usability
\end{itemize}
\end{multicols}
However, with a broader scope, even more indicators should be included, such as:
\begin{itemize}
\item Benevolence (as used in the social sciences)
\item Functional/non-functional capabilities
\item Intervenability
\end{itemize}
In technical terms, we would rather call these indicators ``trustworthiness characteristics'' (TCs) as in \cite{isoiec_trustwhorthiness_2022}. Note that the lists given above are by no means exhaustive. They were selected putting a focus on security and privacy-related TCs.

Moreover, the wide adoption of (in this sense, ``new'') technologies such as artificial intelligence (AI), machine learning (ML), or blockchains and the related distributed ledger technology (DLT) might require an extension of the set of TCs and will influence the fundamental trustworthiness assumptions with regard to a given system. The current ``black box'' nature of many AI/ML algorithms leads to a situation where explainability emerges as a fundamental precondition for defining AI/ML-related TCs and, hence, ultimately for trustworthiness. Sophisticated explainability could increase the confidence in the correctness of the AI/ML results in the first place; however, it is absolutely necessary when it comes to defining and measuring TCs. An example related to large language models (LLM), such as ChatGPT, would be the output of references supporting the statements made by the LLM. A completely different technology, DLT, allows a system designer to adapt the necessary trust and trustworthiness assumptions: for example, a DLT substituting a central data store shifts the trust from a single trusted entity towards a non-malicious majority of DLT providers. 

In general, a TC is a specific characteristic of a system having specific objectives. A TC is measured with the help of TC measurements, while TC controls are used to implement a given TC.  

The first challenge in reasoning about trustworthiness is to have a comprehensive list of TCs. The second challenge is to define the means of measurability of the TCs. For most of the TCs, there exists no generally accepted means for their measurement. For some TCs (e.g., confidentiality or availability) the measurability appears to be more tangible than for others (like privacy or benevolence). In many cases, a combination of multiple measures is necessary to obtain a comprehensive understanding of the trustworthiness at stake. Depending on the domain and the context in which trustworthiness is being evaluated, the individual weighting of the specific TCs and the means of measurement certainly vary. 

\section{Measurability of Trustworthiness\label{mark-C.}}
Ideally, trustworthiness as a whole should be measurable and quantifiable to allow for comparing systems and to meet the requirement of verifiability of user expectations. There exist various proposals in the literature to create a metric for trustworthiness (e.g., \cite{cho_stram_2019} or the work of ITU-T on ``Trustworthiness evaluation for autonomous
networks including IMT-2020 and beyond''). Yet there is still no holistic trustworthiness metric that can be applied universally. The challenges behind designing a metric for trustworthiness are manifold. One reason is that trustworthiness involves several different TCs \textendash{} and many of them lack sound metrics. Take, for instance, privacy as one TC. Privacy itself comprises different aspects, such as anonymity and unlinkability. However, even for these more basic properties of privacy, no universal applicable metrics exist besides isolated measures like $k$-anonymity or differential privacy \cite{wagner_technical_2018}. 

Another challenge is related to the intended nature of a metric for trustworthiness. Should it be a single number allowing for easy comparison but limiting expressiveness? Should it be a vector containing a value for each TC? This would enhance the expressiveness but makes comparison more difficult. Taking consumer reports as an example, we can often see a combination of both approaches: a single value (e.g., inspired by school marks) for easy and quick assessments, but, additionally, measurements for individual properties of the service or product in question, allowing a consumer to adapt the ``base assessments'' according to their individual situation, beliefs, and requirements. Similar approaches can also be seen in the IT security domain. Here, the Common Vulnerability Scoring System (CVSS), on the one hand, provides a single score, allowing system administrators to do a quick risk assessment. But, on the other hand, it also allows for individual adaptation based on individual requirements (e.g., which protection goals are important) and the individual situation (e.g., do countermeasures exist like firewalls, which mitigate exploitability). Would it, therefore, make sense to define a trustworthiness vector based on the measured TCs and give a weighted trustworthiness score?

Another question regarding a sound trustworthiness metric is whether the trustworthiness value should be time-dependent or time-invariant. A time-dependent metric might be more intuitive in the first place since the assessment of whether, e.g., a cryptographic algorithm can be considered trustworthy might change over time when new weaknesses of the algorithm are discovered (thanks to cryptanalysis), or due to new technological developments (e.g., quantum computers). However, a time-dependent metric would be rather useless if the related values were valid only in the very moment they were measured (or calculated), much less help guide humans in their trust decisions. 

A trustworthiness metric needs to be designed in a way that leads to values which have a reasonable period of validity. Consequently, a metric which leads to time-independent values (or at least to values which equally change over time for similar systems) would be desired. Taking a second look at the example of the cryptographic algorithm and, specifically, the symmetric encryption algorithm DES, one could argue that the inherent security ``level'' provided by DES in terms of effort necessary to find the secret key (through calculations) has not changed substantially since its introduction in the 1970s. What changes is the ``perceived'' or relative security, given the technological advancements in computation power allowing (investing the same amount of money) to break DES today within minutes, while in the 1970s it took years. Translating this to another domain: What was considered a ``fast'' car in the 1930s might be considered a ``slow'' car today \textendash{} although the maximum speed of an individual car is time-invariant.

Given the observations above, a possible solution could be to have two kinds of trustworthiness metrics: one ``absolute'' trustworthiness value, which is time-invariant to reflect the essence of trustworthiness provided by the system and to allow for comparison, and a ``relative'' trustworthiness value, which is time-dependent but captures better the current level of trustworthiness to more easily support human trust decisions. Besides, the need for the latter can even be derived from the definition of trustworthiness itself, which is the ``ability to meet stakeholders’~expectations''. Therefore, the ``level of trustworthiness'' can always be understood as relative to the expectation of a given user (stakeholder).

By reviewing the list of the different TCs in the previous section, it becomes apparent that the TCs are not necessarily independent of each other. This poses additional challenges in optimizing the overall trustworthiness of a system. Consider, for example, TCs which are directly related to protection goals from the IT security domain, e.g., confidentiality, integrity, accountability, availability, anonymity, etc. There are certain dependencies among these protection goals \cite{wolf_properties_2000}. For example, confidentiality and availability weaken each other. Another example: Given the current state of the art, security enhancements often have a negative impact on usability; increased privacy could lead to reduced functionality expressed by the ``privacy vs. utility'' trade-off during system design. Consequently, for an informed trust decision that we can rely on, the correlations of the TCs must be analyzed and taken into account. It is important to understand specific definitions and measuring methods of the TCs to ensure the filtering and weighting of the TCs result in a meaningful trust decision that confirms the learnings and minimizes risks.

\subsection{Related Work\label{mark-D.}}

Recently, the Next Generation Mobile Networks Alliance (NGMN) has published a report on ``6G Trustworthiness Considerations'' \cite{ngm_alliance_6g_2023}. This report emphasizes the need for implementing the following TCs in 6G networks: confidentiality, integrity, availability, privacy, reliability, resilience, and safety. From these needs certain requirements for the design of the upcoming 6G networks are derived. Overall, the report puts a strong emphasis on security and privacy-related aspects of trustworthiness. Moreover, the report does not really differentiate between the concepts of ``trustworthiness'' and ``trust''.

Similarly, Veith et al. published a survey paper analyzing potential trust anchor technologies for 6G systems \cite{veith_road_2023}. Besides technologies, they also analyzed the relationship between technology and trust concepts as developed by humanities. Thereby they also refer to the concepts of ability, integrity, and benevolence. They also identified the question of quantifying and measuring trustworthiness as an important yet unanswered one. As in the NGMN report, no clear distinction between trustworthiness and trust is made. 

Bauer~\cite{bauer} analyzes the relation between trust and trustworthiness from a social science perspective. He agrees with our view of the nature of trust being a subjective one. Our approach is different insofar as we regard trustworthiness an objective characteristic of a technical system.

\section{A Model of Trust and Trustworthiness in the relationship between Humans and Technology\label{mark-IV.}}
Based on the concepts of trust and trustworthiness introduced in section II and measurable TCs in section III, we now develop a unifying model which integrates both concepts. We start with a basic model, which we will extend in the next section.

\subsection{Basic Model integrating Trust and Trustworthiness\label{mark-A2}}
Although TCs can be measured in principle, above all, ``absolute'' measurements of TCs must be put into context and are only a prerequisite for making a trust decision. Additionally, emotions play a role in human decisions, which, in technical terms, are related to a combination of general long-term and acute experiences (i.e., situation context). 

Hence, the selection and weighting of the objective measurable TCs in the decision-making process depends on context and experience. In addition, the decision-making is based on the accumulated experiences collected from previous decisions and actions as well as the current experience (emotional state), as shown by empirical studies (e.g.,~\cite{roleTrust}). This forms a feedback loop of experiences that needs to be modelled as well. It depends on a couple of key factors that play a role in making a trust decision. These key factors are:
\begin{enumerate}
\item Available TCs (objective)
\item Personal experience \& learnings (subjective)
\item Current state (subjective)
\end{enumerate}
\begin{figure}[b!]
    \centering
    \includegraphics[width=7.35cm]{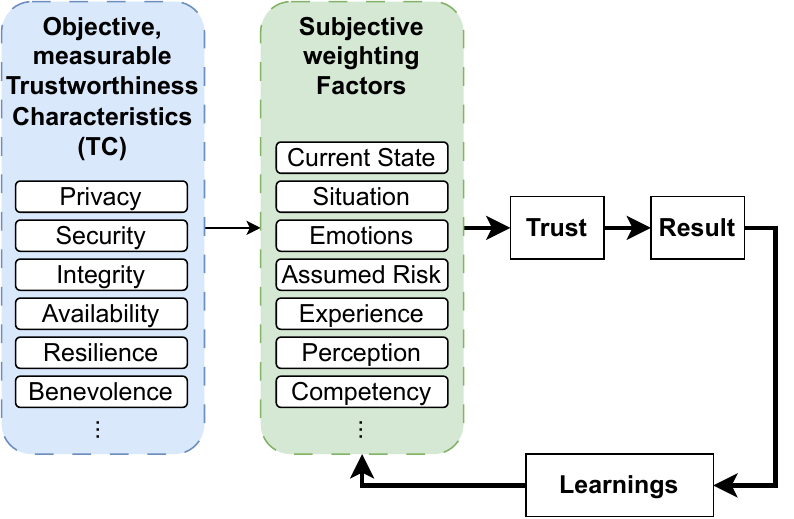}
    \caption{Basic trust model of human-technology interactions}
    \label{fig:BasicModel}
\end{figure}

In an ideal situation, humans should take objective measurements of trustworthiness characteristics (TCs), the current situation (one's own state), as well as their learnings into account to make an informed decision on what to trust and to which extent. Eventually, the observed experience based on the decision will be added to the memory of learnings. Thus, the available TCs should be taken into account, filtered, and weighted depending both on the state and on the memory of learnings. In turn, the learnings depend on the observations and experiences from previous similar situations. Fig.~\ref{fig:BasicModel} shows this basic feedback loop as a block diagram. 

Note that, in reality, the trust decisions of a human being might only slightly be influenced by objective TC measurements. Indeed, people might just end up deciding on the green car instead of the blue one, for example, as a result of their subjective weighting  \textendash{} not only because they like green more than blue but because the color might emotionally lead to trust and trust-related decisions. Remember that according to Mayer et al., trust-based decisions are related to dealing with risks for the trusting party. Obviously, the color of the car will only have negligible influence on the safety of the car, and, therefore, the foundation of such a trust decision might not contribute to the intended risk reduction \textendash{} even if it might contribute to a reduction of the \textit{perceived} risk. Therefore, it is of utmost importance that systems which can harm humans are designed and operated following the ``trustworthiness by design/default'' paradigm. In this way, humans can indeed make subjective trust decisions and still will not face unacceptable risks, thanks to the safeguards achieved through the objective TCs. Thus, in a proper design, certain TCs must not be left to human weighting. This opens up a societal and regulatory dimension.

\subsection{Extended model: Mediators for dealing with complexity\label{mark-B2}}
\begin{figure}[b!]
    \centering
    \includegraphics[width=\linewidth]{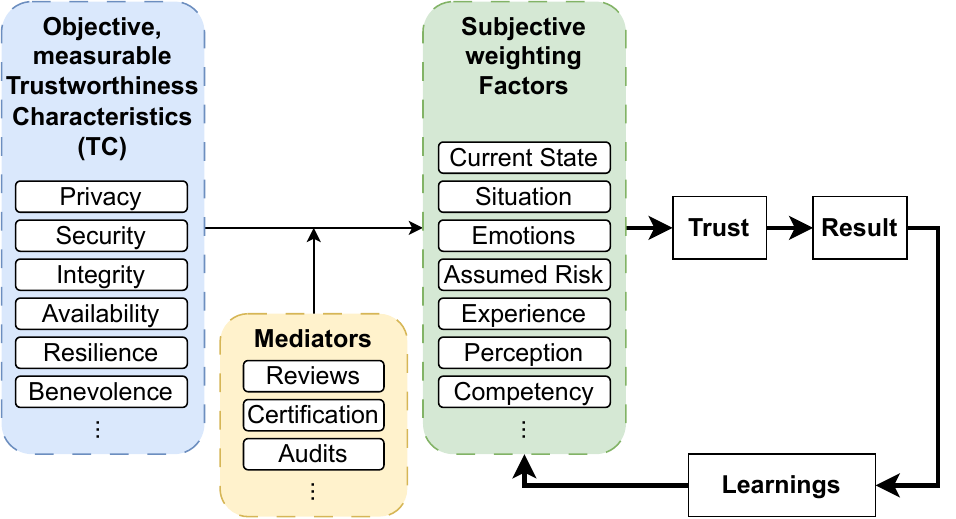}
    \caption{Extended trust model of human-technology interactions}
    \label{fig:ExtendedModel}
\end{figure}

Making trust decisions based on TCs poses the challenge of measuring them and putting them in relation to one's own expectations. People might not be capable or willing to do this on their own. Instead, such measurements can be delegated to third parties in the form of reviews, certification, or audits. We add these third parties as mediators in our model (Fig.~\ref{fig:ExtendedModel}).

The concept of mediators induces a recursive challenge to trustworthiness: The (organizational) trustworthiness of the mediator itself, e.g., of a certification body, influences the (rational) trustworthiness of the result of mediator's concrete action (e.g., certification). On the one hand, we can apply well-known organizational and regulatory means to lower the risk of potentially misbehaving mediators, thus shifting the problem from a purely technical one to a more social sciences-related one. On the other hand, the design of the technical system should support the trustworthiness assessment by every human in the best possible way. Although we stated that most people will not have the ability to do this assessment, some will and can act as an additional safeguard with regard to misbehaving mediators.

Related to the concept of mediators is the concept of reputation. Here, basically, other humans act as mediators, giving their judgement concerning trustworthiness and their past trust-related experiences.
Imagine, for example, a scenario from the domain of autonomous vehicles used in public transportation. Past users can make a judgement based on their experience of using a given autonomous vehicle. The ensuing reputation can then be taken into account by future users when facing trust decisions. 

To support the role of mediators as well as the concept of reputation, trustworthy technical systems must fulfill two fundamental requirements (from which subsequent requirements can be derived). We call them meta-requirements, as they are on a completely different, much more abstract, level than any of the functional or non-functional requirements we discussed so far. The first requirement is the support of auditability, as needed by the mediators, and the second is the support/integration of reputation systems.

\section{Thoughts on an Empirical Evaluation of the Proposed Model}
We derived our trust model from the definition of trustworthiness and the existing ABI model. In the following, we want to justify it by finding empirical evidence that a change in the metrics of the objective TCs indeed influences the trust decisions of humans. A more specific example would be to show that an increase in the measured values of security- and safety-related TCs indeed increases the likelihood that the trust decisions of humans are geared towards systems with higher objective TC values. We selected security- and safety-related TCs as examples because we assume that, if all other things are equal, humans prefer systems which offer higher security and safety. On the other hand, our model implies that not all humans will automatically decide for the system which provides higher security and safety -- because the subjective weighting factors influence the final outcome of the trust decision. 

For example, the study by Wiedemann et al.~\cite{sar} supports this approach. It analyzes the influence of the specific absorption rate (SAR) with regard to customer purchase decisions for mobile phones. Here, we can understand the SAR value as one objective TC, and the purchase decision as the trust decision made by humans. The study results showed that customers perceive higher SAR values as a greater risk, leading to purchasing decisions towards phones with lower SAR values. This can be interpreted as an objective TC which influences the trust decisions of the customers. On the other hand, the study showed that a lower SAR value of a given phone does not automatically lead to a purchase decision for that phone -- this could be seen as an indicator that subjective weighting factors eventually influence the trust decision. 

Other examples are the various battery and overheating problems of mobile phones, as with the Motorola Razr (2005), HTC One M9 (2015) or Samsung Galaxy Note 7 (2016). In all cases, these problems led to a decrease in sales, which again can be seen as some evidence for the claim that a change (i.e., decrease) in an objective (safety-related) TC influences trust-based decisions (i.e., sales). 
These examples also show the importance of the mediators since, in fact, the objective TC had not changed -- only the knowledge about it, thanks to reports in the media. 

On the other hand, there are cases in which low metrics of certain objective TCs become known, and such knowledge does not lead to changing trust decisions, nor, specifically, to decreases in sales either. We could attribute this behavior of the consumers to the subjective weighting factors of our model. One example is the iPhone 6 ``Bendgate'' (2014), which came to be known to be susceptible to chassis bending. Nevertheless, the iPhone 6 achieved sales records. A second example is the HTC One M8 overheating issue (2014), which did not lead to notable decreases in sales. 

A third example can be taken from the vehicular domain: Adaptive cruise control (ACC) systems enable a vehicle to automatically keep a minimum distance to vehicles in front. From a technical, i.e., safety, related point of view, it would be sufficient to have a fixed, predefined minimum distance (and which is potentially speed dependent). Nevertheless, many car manufacturers allow users to configure a larger distance than necessary. In this example, the trust-related decision is to enable/use the ACC system. Given a certain type of car, the objective TCs related to the ACC system are basically equal for all users. Therefore, the option to configure the minimum distance accounts for the subjective weighting factors of individual drivers. 

\section{Conclusion\label{mark-VI.}}
In this paper, we presented a model which strives to clarify the relationship between ``trustworthiness'' and ``trust'' in the interaction of humans with technology. It is important to understand that trustworthiness and trust are not synonymous but have fundamentally different meanings. Trustworthiness refers to \textit{objective characteristics} of a given technical system, whereas trust relates to \textit{subjective behavior or beliefs} of humans. Ideally, trustworthiness influences the trust decisions in a way which reduces the risks for human beings. This concept could possibly be extended to machine-to-machine interactions.

Thorough reasoning about trustworthiness imposes many challenges. One of them is the lack of sound means for measurability regarding the many different trustworthiness characteristics as well as meaningful methods to combine these measurements into an overall trustworthiness measurement.

Nevertheless, a trustworthy system design is of utmost importance \textendash{} especially regarding the upcoming 6G communications system that will connect an unheard-of number of physical objects we deal with every day. This challenge must be approached in a holistic manner supporting auditability and reputation systems.

The paradigm of ``trustworthiness by design/default'' does not just call for security technologies to be incorporated. If truly embraced, it will change system design very fundamentally.

\section*{Acknowledgements}
We thank our partners from the Disruption and Societal Change Center (TUDiSC) of TU Dresden for the fruitful knowledge exchange. We especially thank the ``Digital disruption and disinformation: challenges to institutional legitimacy and trustworthiness'' (DiDis) project -- namely Marianne Kneuer, Sophie Noack, and Johanna Wolter -- and the ``Disruptions of networked privacy'' (DIPCY) project.

The authors are financed based on the budget passed by the Saxon State Parliament in Germany. This work has been partly funded by the European Commission (project Hexa-X-II, Grant no. 101095759) and by the Federal Ministry of Education and Research, Germany (Grant no. 16KISK046).
\bibliographystyle{plain}
\bibliography{references}

\begin{IEEEbiographynophoto}{Gerhard P. Fettweis}
(Fellow, IEEE) received his PhD under H. Meyr at RWTH Aachen in 1990, and thereafter worked at IBM and TCSI in California. Since 1994, he has been the Vodafone Chair Professor for Mobile Communications Systems at TU Dresden. Since 2018, he has also been the founding scientific director and CEO of the Barkhausen Institut. His research focuses on physical layer design and its hardware implementation. His team has spun out 19 startups. He is a member of the German Academy of Sciences (Leopoldina), the German Academy of Engineering (acatech), and the National Academy of Engineering (NAE).
\end{IEEEbiographynophoto}

\begin{IEEEbiographynophoto}{Patricia Grünberg}
 researched the topic of trust and trustworthiness in the interdisciplinary context of social and economic sciences. She currently heads the Administration and Public Outreach groups at the Barkhausen Institut. 
\end{IEEEbiographynophoto}

\begin{IEEEbiographynophoto}{Tim Hentschel}
 received his PhD in digital signal processing for communications 
from TU Dresden in 2001. He co-founded a start-up and held 
several engineering management positions thereafter. Since 2018, he has been 
managing the scientific and engineering research operations of the Barkhausen 
Institut.
\end{IEEEbiographynophoto}

\begin{IEEEbiographynophoto}{Stefan Köpsell}
 has been working in the field of trustworthiness in and by distributed systems for more than 25 years. He currently heads the ``Trustworthy Data Processing'' group at the Barkhausen Institut. He is also a member of the Chair of Privacy and Security at TU Dresden. His research focuses on privacy-enhancing technologies in the area of the Internet of Things, with a strong focus on next-generation mobile networks.
\end{IEEEbiographynophoto}
\vfill
\end{document}